\documentclass[a4paper]{ESASPCS13Style}
\usepackage{epsfig}

\begin{document}

\title{A VLT/UVES spectroscopic analysis of C-rich Fe-poor stars}

\author{T. Masseron\inst{1,2}, B. Plez\inst{2}, F. Primas\inst{1}, 
S. Van Eck\inst{3}, \and A. Jorissen\inst{3}} 
\institute{ESO, Garching, Germany, 
           \and GRAAL, Universit\'e Montpellier II, France,
  \and Institut d'Astronomie et d'Astrophysique, 
  Universit\'e Libre de Bruxelles, Belgium}

\maketitle 

\begin{abstract}
Large surveys of very metal-poor stars have revealed in 
recent years that a large fraction of these objects were carbon-rich,
analogous to the more metal-rich CH-stars. 
The abundance peculiarities of CH-stars are 
commonly explained by 
mass-transfer from a more evolved companion.  In an effort to better
understand the origin and importance for Galactic evolution of
Fe-poor, C-rich stars,
we present abundances determined from high-resolution and high 
signal-to-noise spectra 
obtained with the UVES instrument attached to the ESO/VLT. 
Our analysis of carbon-enhanced objects includes 
both CH stars and more metal-poor objects, and we explore the 
link between the two classes.  
We also present preliminary results of our ongoing radial 
velocity monitoring.

\keywords{Stars: abundances, Stars: carbon, Stars: Population II, 
Stars: binaries: general }
\end{abstract}

\section{Introduction}
A subgroup of carbon stars, the CH stars, were first 
distinguished
from other carbon stars 60 years ago (\cite{keenan42}). 
Their chief characteristics
are : strong CH absorption lines, strong Swan bands
of C$_2$, strong resonance lines of Ba~II and Sr~II
(later shown to reflect genuine overabundances of 
s-process elements),
weak lines of the iron-group elements and high proper motion.

 Since it was shown that all field CH stars
 were spectroscopic binaries (\cite{mcclure90}),
 the binary scenario was accepted: CH stars have been 
 polluted by a nearby
companion, formerly on the AGB, now a defunct white dwarf.

 But some recent results have cast doubts on this 
 general picture.
 Conflicting evidences include the radial velocity
  monitoring of targets from the HK survey (\cite{beers92}) 
  which show no radial velocity variations (\cite{preston01}), 
  and the discovery of carbon but not s-process enhanced stars 
  (e.g. \cite{sneden96})
 . Furthermore, the unexpectedly high fraction of 
 carbon-enhanced stars
among metal-poor stars discovered in the HK survey 
(about 25\% in the metallicity range [Fe/H] $\le$ -2.5 compared 
to 1-2 \% among stars of higher metal abundances), make 
them crucial for the study of the early stages of Galaxy evolution.

\begin{table}
\caption{The sample and the atmospheric parameters.}
\label{table}
\begin{center}
\leavevmode
\footnotesize
\begin{tabular}{l|c|c|c|c} 
\hline
star     &T$_{\rm eff}$  & log~g &  [Fe/H]  & $^{12}$C/$^{13}$C  \\
\hline
CS 22891-171      &     5100  &  1.8  &  -2.31   &  5     \\ 
CS 22942-019      &     5100  &  2.5  &  -2.48   &  7     \\ 
CS 22945-017*    &     6400  &  4.3  &  -2.60   &  4     \\ 
CS 22953-003      &     4600  &  1.0  &  -3.27   & $>$5   \\ 
CS 22956-028*    &     6700  &  3.5  &  -2.38   &  3     \\ 
CS 30322-023      &     4000  &  0.0  &  -3.78   &  5     \\ 
HD 26*                &     5200  &  2.6  &  -0.71   &  ?     \\
HD 5424*            &     5000  &  3.0  &  -0.41   &  4    \\     
HD 24035*          &     5000  &  3.7  &   0.16   &  4     \\   
HD 168214          &     5200  &  3.5  &  -0.03   &  ?     \\ 
HD 187861*        &     4800  &  1.8  &  -2.21   & 10    \\  
HD 196944*        &     5250  &  1.7  &  -2.21   &  ?     \\ 
HD 206983          &     4550  &  1.6  &  -0.93   &  4     \\ 
HD 207585*        &     5800  &  4.0  &  -0.35   & 10    \\   
HD 211173          &     4800  &  2.5  &  -0.17   & 13    \\   
HD 218875          &     4600  &  1.5  &  -0.63   & 100   \\ 
HD 219116          &     4800  &  1.8  &  -0.45   &  7      \\ 
HD 224959*        &     5000  &  2.0  &  -2.08   &  7     \\  
HE 1419-1324     &     4900  &  1.8  &  -3.28   &  4      \\ 
HE 1410+0213* &     4550  &  1.0  &  -2.38   &  4      \\ 
HE 1001-0243*   &     5000  &  2.3  &  -3.12   & 30     \\ 
\hline
\multicolumn{5}{l}{* probably binary, from radial velocity monitoring}\\
\end{tabular}
\end{center}
\end{table}

\section{Observations and abundance determinations}
We have started a new extensive high-resolution analysis of 
carbon-enhanced stars, based on a sample including at the 
moment 86 stars, aimed at investigating the C-enrichment 
phenomenon over a wide range of metallicities, chosen in the 
Barktevicius catalog (1996), in the HK survey and in the 
Hamburg/ESO survey (\cite{christlieb01}). Observations 
were made with the VLT/UVES echelle spectrograph 
at high S/N and spectral resolution, as well as with the 
ESO 1m52/FEROS spectrograph. A few radial velocity observations were made
with the OHP 1m93 ELODIE instrument.\\
In Table~\ref{table}, we present a subset of our sample for 
which we have derived C, N (from CN and CH bands), O 
(from [O~I] 630nm, when possible), 
and Ba and Eu abundances, from VLT observations.\\

Effective temperatures were initially estimated from photometry 
available in the literature, using the 
calibration of \cite{alonso99}. However, due to strong \
absorption by molecular CH and CN bands impacting the photometry
by an unknown amount, T$_{\rm eff}$ were instead
derived by forcing the abundance determined 
from individual Fe I lines to show no dependence on excitation
 potential. The gravity was determined
 from the ionization equilibrium  of Fe~I and Fe~II. The microturbulent 
 velocity was set by requiring 
 no trend of Fe~I abundance with equivalent width.
The observed spectra are compared to synthetic ones, computed 
with the "turbospectrum" package
 (\cite{alvaplez98}). This program uses OSMARCS atmosphere 
 models, initially developed by \cite{gustafsson75} and 
 later improved by \cite{plez92}; see \cite{gustafsson03}
  for details on recent improvements. \\
The line lists are the same as used by \cite{hill02} for 
atoms and for  CH, C$_2$ and CN, and their various isotopes.\\
\begin{figure}
\begin{center}
\epsfig{file=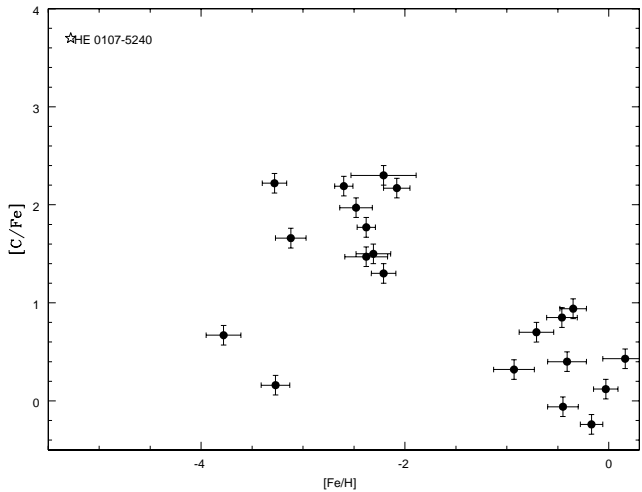, width=7cm, angle=-90}
\end{center}
\caption{Carbon abundance vs. iron abundance. Despite some 
scatter, an increase of the carbon overabundance with decreasing
Fe/H is clearly visible. Note that the most Fe-poor star 
HE~0107-5240 (Christlieb et al. 2002) follows the trend.\label{figcarbon}}
\end{figure}

\section{Results}
\begin{figure}
\begin{center}
\epsfig{file=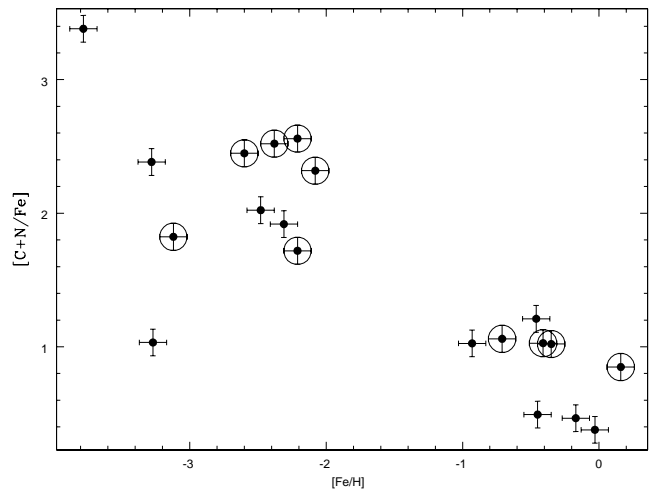, width=7cm, angle=-90}
\end{center}
\caption{Carbon plus nitrogen abundance vs. iron abundance.
For the majority of the stars, the global enrichment in
(C+N)/H is approximately independent of Fe/H, indicating a 
primary origin of C (and N, possibly through posterior 
CN-processing). Suspected binaries, detected through radial 
velocity variations, are circled.
\label{figcarbonnitrogen}}
\end{figure}

\begin{figure}
\begin{center}
\epsfig{file=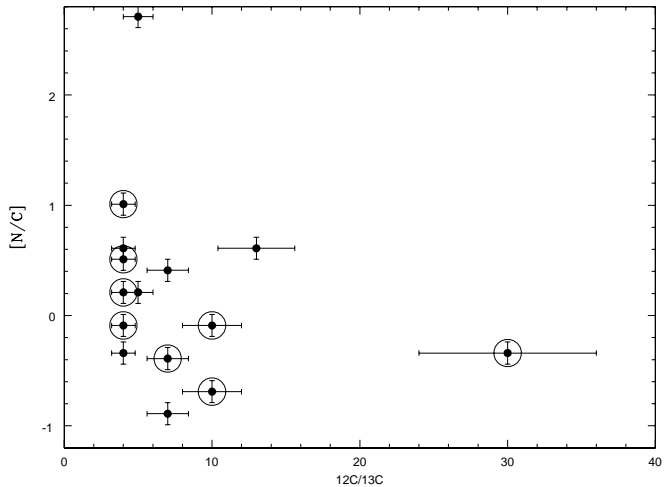, width=7cm, angle=-90}
\end{center}
\caption{Nitrogen enrichment vs. $^{12}$C/$^{13}$C ratio, showing
the operation of the CN-cycle. Suspected 
binaries are circled.\label{fignitrogen}}
\end{figure}

\begin{figure}
\begin{center}
\epsfig{file=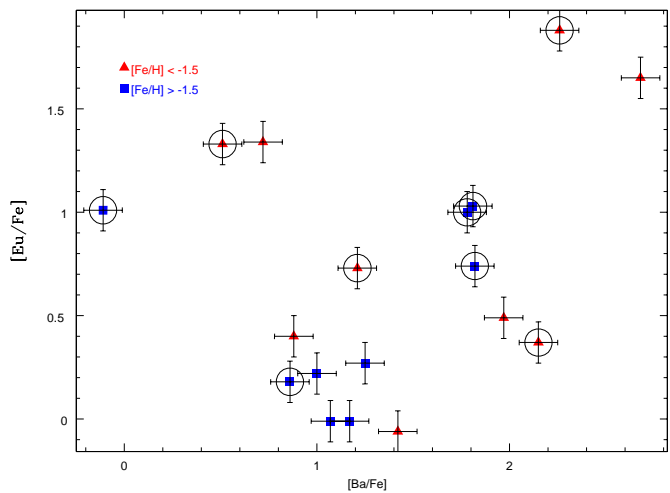, width=7cm, angle=-90}
\end{center}
\caption{Eu vs. Ba abundances. The various origin of carbon-rich,
Fe-poor stars is obvious in this diagram, where stars enriched
in r-process elements, or in s-process, or maybe both, seem
to exist at various metallicities.Suspected binaries are circled.
\label{figeuba}}
\end{figure}

Figures~\ref{figcarbon}, \ref{figcarbonnitrogen}, \ref{fignitrogen}, 
and \ref{figeuba}
 present  the derived abundances. 
Carbon is indeed enhanced in the atmospheres of most of our 
metal-poor targets (Fig.~\ref{figcarbon}). 
Excluding the 2 stars around [Fe/H] $= -3.5$ that do not show 
a large C enhancement, the average carbon abundance is almost 
constant  from the CH stars ([Fe/H] $> -2.0$) 
to the very metal-poor carbon-enhanced stars 
(from [C/H]$\approx 0$ at solar Fe/H to [C/H]$\approx -0.5$
at [Fe/H]$\approx -3$). 
The lowest metallicity star HE 0107-5240 ([Fe/H] $=-5.3$, recently 
discovered by \cite{christlieb02}), shows a carbon abundance 
close to the solar value. This extreme star follows the trend of 
Fig.~\ref{figcarbon}.
The nitrogen abundance, when combined to the carbon abundance
and the carbon isotopic ratio, provides additional information.
Figure~\ref{figcarbonnitrogen} shows the sum of C and N overabundance
as a function of Fe/H. Some of the stars that did not show a large C/Fe
do show a large (C+N)/Fe, as the most metal-poor star of our sample.
The $^{12}$C/$^{13}$C ratio is generally low (see Table~\ref{table}), 
often close to the 
CN-cycle equilibrium value.
There is a general anticorrelation between the N overabundance and 
$^{12}$C/$^{13}$C (Fig.~\ref{fignitrogen}), characteristic
of the operation of the CN cycle.
Note that the stars of our sample with low $^{12}$C/$^{13}$C
and still on the main 
sequence (cf. log~g values in Table~\ref{table}) are binaries, 
supporting the mass-transfer scenario.
The carbon and nitrogen enhancements are large, and whether
the CN cycling happened in the stars we observe when they evolved 
to the giant stage, or in the companions that polluted them,
remains to be determined. The combined (C+N)/Fe overabundance is 
much larger at lower metallicity, pointing towards a primary origin of
these elements.
\cite{asplund03} warns for 3D effects that may 
lead to overestimates of abundances derived from molecular lines 
in metal-poor stars, but we doubt that the trend could be 
erased by NLTE and 3D effects.\\

Ba is an s-process element and, as for carbon, its overabundance 
is explained by 
mass-transfer from a now extinct AGB star. The r-process element Eu is 
believed to originate from 
supernovae, although the r-process site(s) is still debated. 
Figure~\ref{figeuba} shows [Eu/Fe] vs. [Ba/Fe] for the stars of 
Table~\ref{table}. 
In addition 
to a  few stars that are very 
enriched in both r- and s-process elements, two groups emerge: 
one enriched in Eu and not in 
Ba, and a large group of stars with [Ba/Fe] around +1, and no 
or little Eu enhancement. 
This latter group encompasses the solar metallicity Ba stars 
present in our sample.\\

The carbon-rich, very iron-poor stars are of various origins: 
(i) mass-transfer binaries polluted by
an AGB or a more massive star, 
(ii) single stars, some enriched in s-process elements, other
in r-process elements, some maybe in both. 
We are pursuing our  
analysis of more stars
in our sample, and of more chemical elements, in order to 
provide useful constraints
on their origin, and on the early chemical evolution of the Galaxy.

\end{document}